\documentclass[useAMS,usenatbib]{mn2e}
\usepackage{graphicx,color,epsfig}

\newcommand{\be}{\begin{equation}}
\newcommand{\ee}{\end{equation}}

\newcommand{\apj}{ApJ}
\newcommand{\icarus}{ICARUS}

\newcommand{\mnras}{MNRAS}
\newcommand{\aap}{A\&A}

\newcommand{\nat}{Nature}

\def\ltsima{$\; \buildrel < \over \sim \;$}
\def\simlt{\lower.5ex\hbox{\ltsima}}
\def\gtsima{$\; \buildrel > \over \sim \;$}
\def\simgt{\lower.5ex\hbox{\gtsima}}

\def\msun{{\,{\rm M}_\odot}}

\newcommand\mearth{{\,{\rm M}_{\oplus}}}

\def\del#1{{}}

\title[Planet rotation and the Moon]{Rotation of the Solar System planets and
  the origin of the Moon in the context of the tidal downsizing hypothesis.}

\author[S. Nayakshin]{Sergei Nayakshin \\ Department of Physics \& Astronomy,
  University of Leicester, Leicester, LE1 7RH, UK}

\begin{document}

\date{Accepted 2008 ?? ??. Received 2008 ?? ??; in original form 2008 05 ??}

\pagerange{\pageref{firstpage}--\pageref{lastpage}} \pubyear{2008}

\maketitle

\label{firstpage}

\begin{abstract}
It has been proposed recently that the first step in the formation of both
rocky and gas giant planets is dust sedimentation into a solid core inside a
gas clump (giant planet embryo). The clumps are then assumed to migrate closer
to the star where their metal poor envelopes are sheared away by the tidal
forces or by an irradiation-driven mass loss. We consider the implications of
this hypothesis for natal rotation rates of both terrestrial and gas giant
planets. It is found that both types of planets may rotate near their break up
angular frequencies at birth. The direction of the spin should coincide with
that of the parent disc and the star, except in cases of embryos that had
close interactions or mergers with other embryos in the past. Furthermore, the
large repository of specific angular momentum at birth also allows formation
of close binary rocky planets inside the same embryos. We compare these
predictions with rotation rates of planets in the Solar System and also
question whether the Earth-Moon pair could have been formed within the same
giant planet embryo.
\end{abstract}


\section{Introduction}\label{intro}

Recently, a ``tidal downsizing'' hypothesis \citep{Nayakshin10c} for planet
formation was advanced \citep[see
  also][]{BoleyEtal10,Nayakshin10a,Nayakshin10b}. Planets, both rocky and gas
giants, are built in this hypothesis very early on, while the gaseous disc is
still comparable in mass with the star. In brief, the disc is expected to
fragment on gaseous clumps with mass $\sim 10$ Jupiter masses at $\sim 100$ AU
scales, where radiative cooling is sufficiently fast. The gas clumps then
contract due to radiative cooling. The contraction process may be protracted
enough \citep{Nayakshin10a} to allow the dust to sediment inside the embryos
to make terrestrial planet cores \citep[as proposed
  by][earlier]{Boss98}. Finally, embryos (gas clumps) migrate closer to the
star, where their gaseous envelopes are tidally and possibly irradiatively
disrupted, leaving behind either rocky cores of terrestrial planets or more
massive gas giants.

Numerical simulations of massive gas discs by e.g., \cite{VB06,BoleyEtal10}
appear to support the embryo migration part of the hypothesis, while the
recent simulation by \cite{ChaNayakshin10} has actually resulted in a
``super-Earth'' solid core being delivered from $\sim 100$ AU to $\sim 8 $
AU. Nevertheless, the numerical simulations of this kind are in their infancy,
and it also remains unclear how robust the results are given that the embryo
evolution strongly depends on assumed dust opacity and other parameters of the
problem.

A supplementary way to test a hypothesis is to consider its least model
dependent predictions and contrast them to observations. In this paper we make
one such comparison by considering the spins of the planets at birth in the
context of the tidal downsizing scheme. We point out that gas clumps born in
the disc by fragmentation are usually found to rotate in prograde direction
with the spin tightly aligned to that of the parent disc. We show below that
rotation of the giant embryos endows both rocky and giant planets born inside
the embryos with prograde rotation at high, potentially near break-up,
rates. The offsets of planet's direction of spin from the disc rotation in
this scenario is due to embryo-embryo interactions. We also note that rapid
rotation of the inner zones of the embryos implies that rocky planets born
there by gravitational instability may not form single but be in binaries or
even multiples.

These predictions are consistent with the observations of the
  Solar System planets rotation pattern, e.g., relatively rapid and mainly
  prograde. We also note that the Earth and the Moon would have to be born
  inside the same giant planet embryo if the tidal downsizing hypothesis is
  correct.

We conclude the paper by noting that despite these encouraging results, there
is a whole list of observations (see \S \ref{sec:other} compiled partly due to
the anonymous referee of this paper) that the tidal downsizing hypothesis
needs to be further tested upon.

\section{Simulated Rotating Giant planet embryos}\label{sec:rotating_embryo}

Although our arguments are analytical, we find it useful to illustrate our
points with numerical simulations that were recently presented by
\cite{ChaNayakshin10}, who simulated fragmentation and evolution of a massive,
$M_d = 0.4 \msun$, gas disc around a parent star of mass $M_* = 0.6
\msun$. The gas component was modelled with a 3D SPH code utilising an
analytical approximation to the radiative cooling, whereas the dust was
treated as a second fluid under the influence of gravity and the drag force
from the gas. The grains were allowed to grow via a stick-and-hit mechanism
saturated at a maximum impact velocity of $3$ m s$^{-1}$.

As expected, the disc fragmented on a dozen or so gaseous clumps with masses
between 5 and 20~$M_J$ at a distance of $\sim 70$ to $\sim$ 150 AU. Some of the clumps
merge with one another, others interact strongly gravitationally. Several
clumps make it into the inner few tens of AU. The less dense ones are
destroyed by tidal shear releasing their dust content at $\sim 15$ AU. One
particular embryo was dense enough to spiral in closer before being completely
destroyed. By virtue of its higher density the embryo also contained larger
dust grains and a gravitationally collapsed dust core of mass $\sim 7.5
\mearth$. The ``super-Earth'' core was deposited in a low eccentricity orbit
with the semi-major axis of $\approx 8$ AU.

\begin{figure}
\centerline{\psfig{file=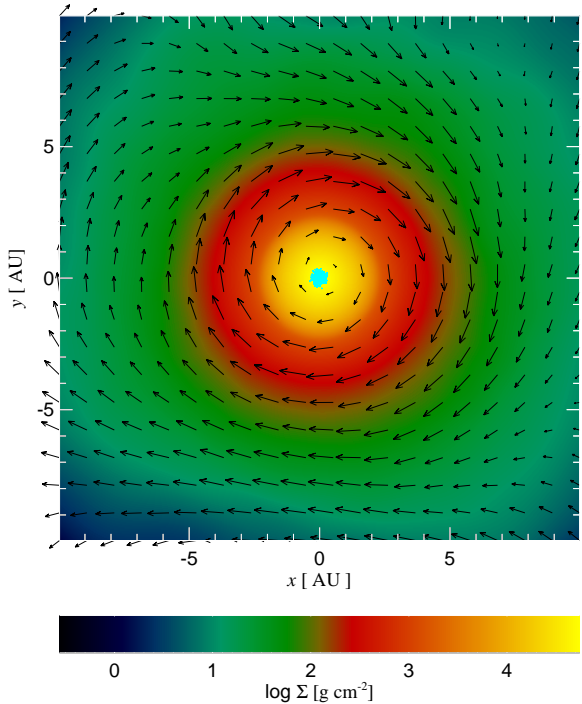,width=0.5\textwidth,angle=0}}
\caption{The top projection of a typical ``undisturbed'' embryo at a large
  ($\sim 100$~AU) distance from the star. The central cyan colour points show
  dust particles with size greater than 10 cm, and vectors
    show the velocity field.}
\label{fig:embryo_disc}
\end{figure}

Figure \ref{fig:embryo_disc} shows the face-on gas column density map centred
on a typical undisturbed embryo at a large distance from the star at time
$t=4880$ years \citep[same as the left panel of figure 4
  in][]{ChaNayakshin10}. The collection of cyan dots in the centre of the
figure is the dust grain particles with size greater than 10 cm. The grain
concentration is not yet high enough to yield a collapsed solid core at the
time of the snapshot.  Black arrows show the velocity field of the gas with
respect to the velocity of the densest part of the embryo. The spin direction
is the same as that of the parent disc around the star, save for offset by
about $5^\circ$. The origin of prograde rotation of the embryo may be in the
shape of streamlines on the ``horse-shoe orbits'' of gas near 
massive planets \citep[see][]{LubowEtal99}.  The magnitude of velocity vectors
in the figure first increase with distance from the centre (to the distance of
a few AU), and then stay roughly constant or perhaps decrease slightly further
out.

\begin{figure*}
\centerline{\psfig{file=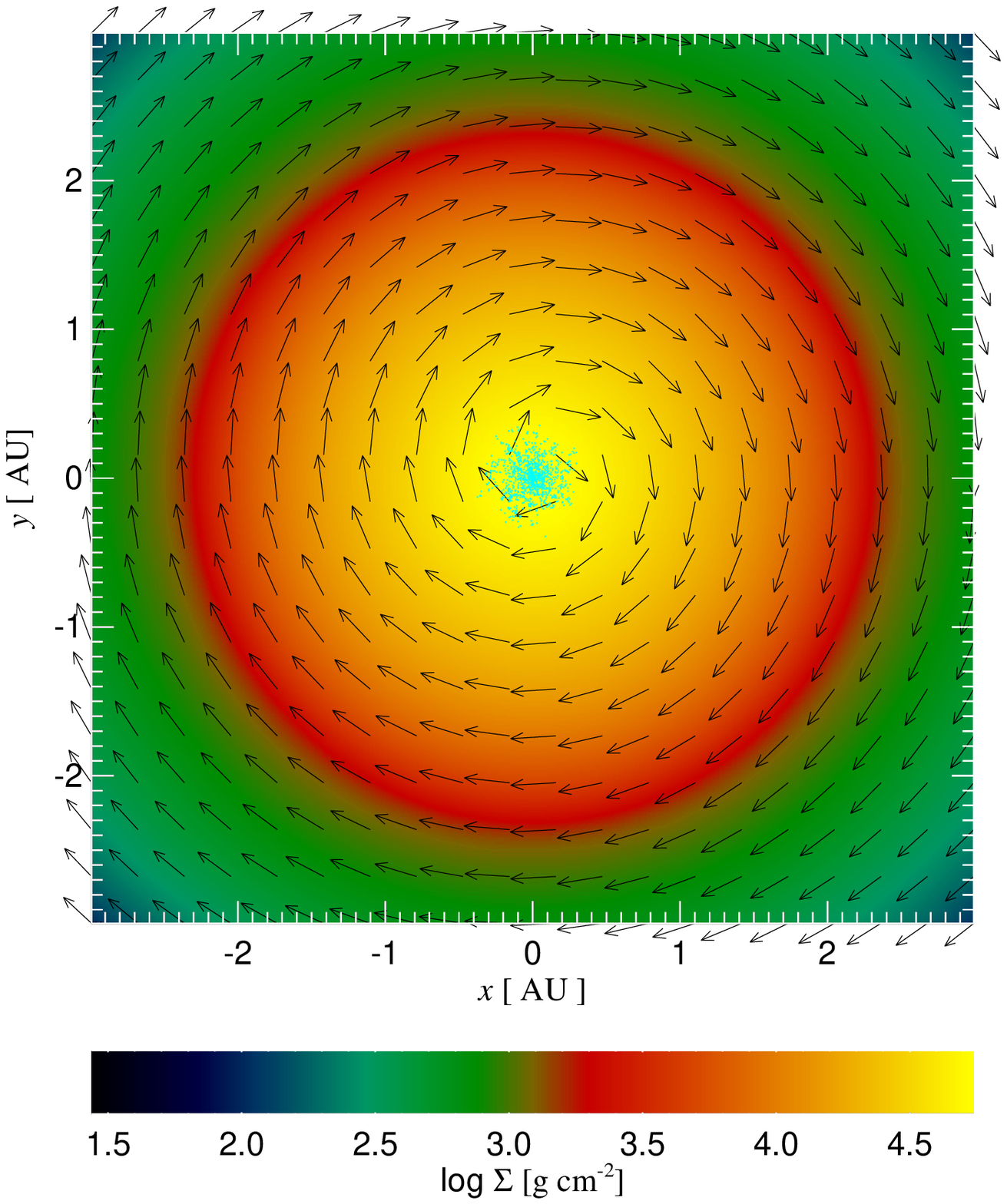,width=0.5\textwidth,angle=0}
\psfig{file=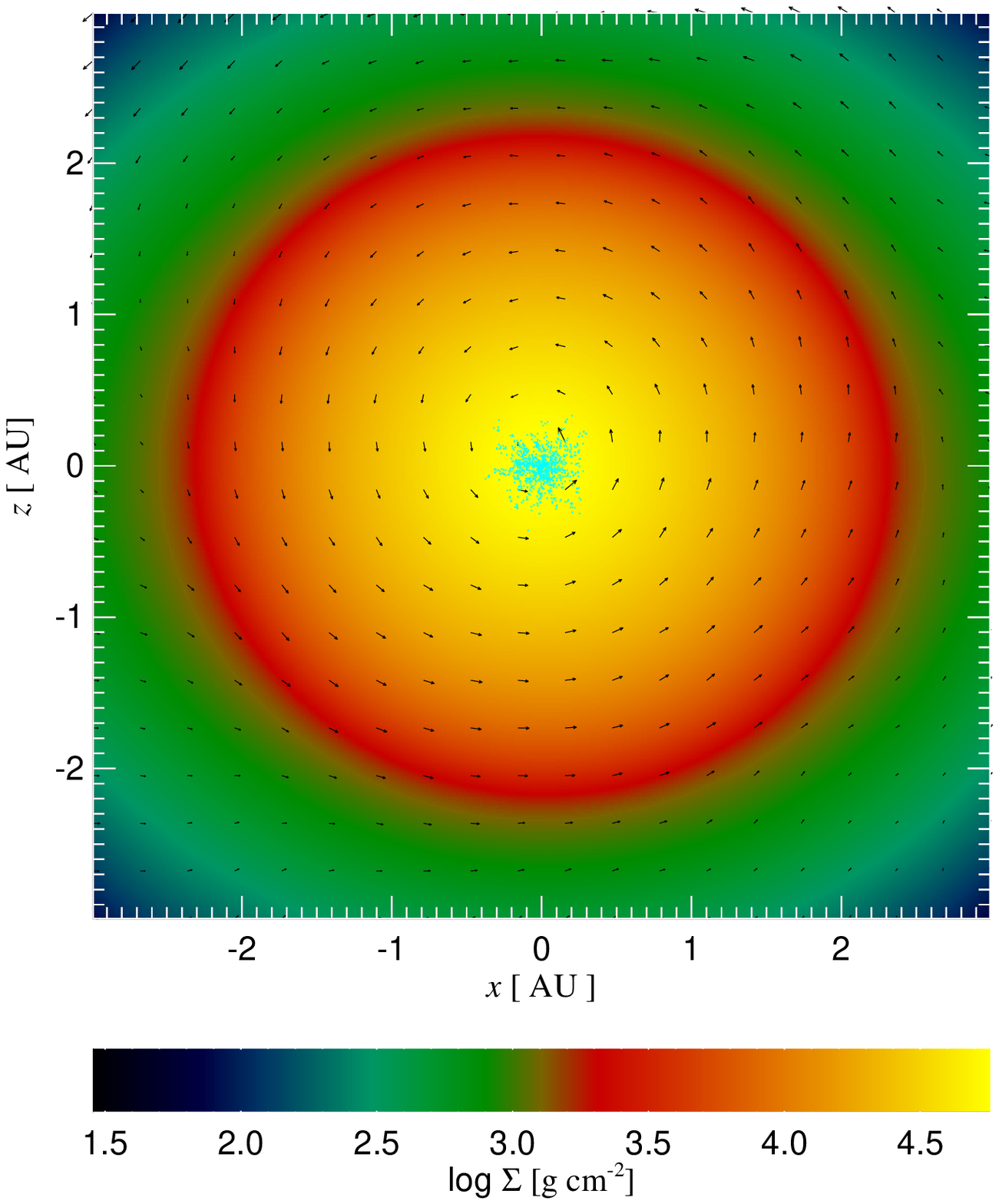,width=0.5\textwidth,angle=0}}
\caption{Left: Same embryo at same time as in Figure \ref{fig:embryo_disc} but
  at smaller scales and using the ``solid body rotation'' normalisation for
  velocity vectors. The near constancy of the magnitude of the velocity
  vectors indicates that the embryo rotates as a solid body. Right: edge-on
  projection of same embryo. Note the flattened Saturn-like shape of the
  envelope. The scaling of velocity vectors is slightly increased for improved
  visibility of the flow directions in this panel.}
\label{fig:rotating_embryo}
\end{figure*}

A sufficiently viscous gaseous body may be expected to rotate at a constant
angular frequency, e.g., as a solid body. For a constant density embryo model
\citep{Nayakshin10a}, the maximum break-up angular frequency of rotation is
\begin{equation}
\Omega_{\rm break} = \left(\frac{4\pi G\rho_0}{3}\right)^{1/2}\;,
\label{omega_break}
\end{equation}
where $\rho_0$ is the density of the embryo. We define the rotational break-up
velocity of the embryo as
\begin{equation}
v_{\rm break} = \Omega_{\rm break} r\;,
\label{v_break}
\end{equation}
where $r = \sqrt{x^2 + y^2}$ is the projected distance to the embryo's
centre. To analyse the rotation pattern of embryo from figure
\ref{fig:embryo_disc} further, we normalise the velocity field on $v_{\rm
  break}$, taking $\rho_0$ to be the mean gas density in the embryo. The left
panel of Figure \ref{fig:rotating_embryo} shows the embryo in the same face-on
projection as in figure \ref{fig:embryo_disc}, whereas the right panel of the
figure shows the projection of the embryo along the $y$-direction (e.g., in
the plane of the disc).  Velocity vectors normalised on
  $v_{\rm break}$ are also plotted.

It is clear from Fig. 2 that the embryo spins nearly as a
solid body in the central few AU, e.g., with a constant
  angular frequency.  Furthermore, the right panel shows that the rotation is
significant enough to deform the embryo's shape from a spherical shape to that
of an oblate spheroid flattened along the spin vector. The amplitude of
rotation $\Omega$ is close to $0.1\Omega_{\rm break}$.

\begin{figure*}
\centerline{\psfig{file=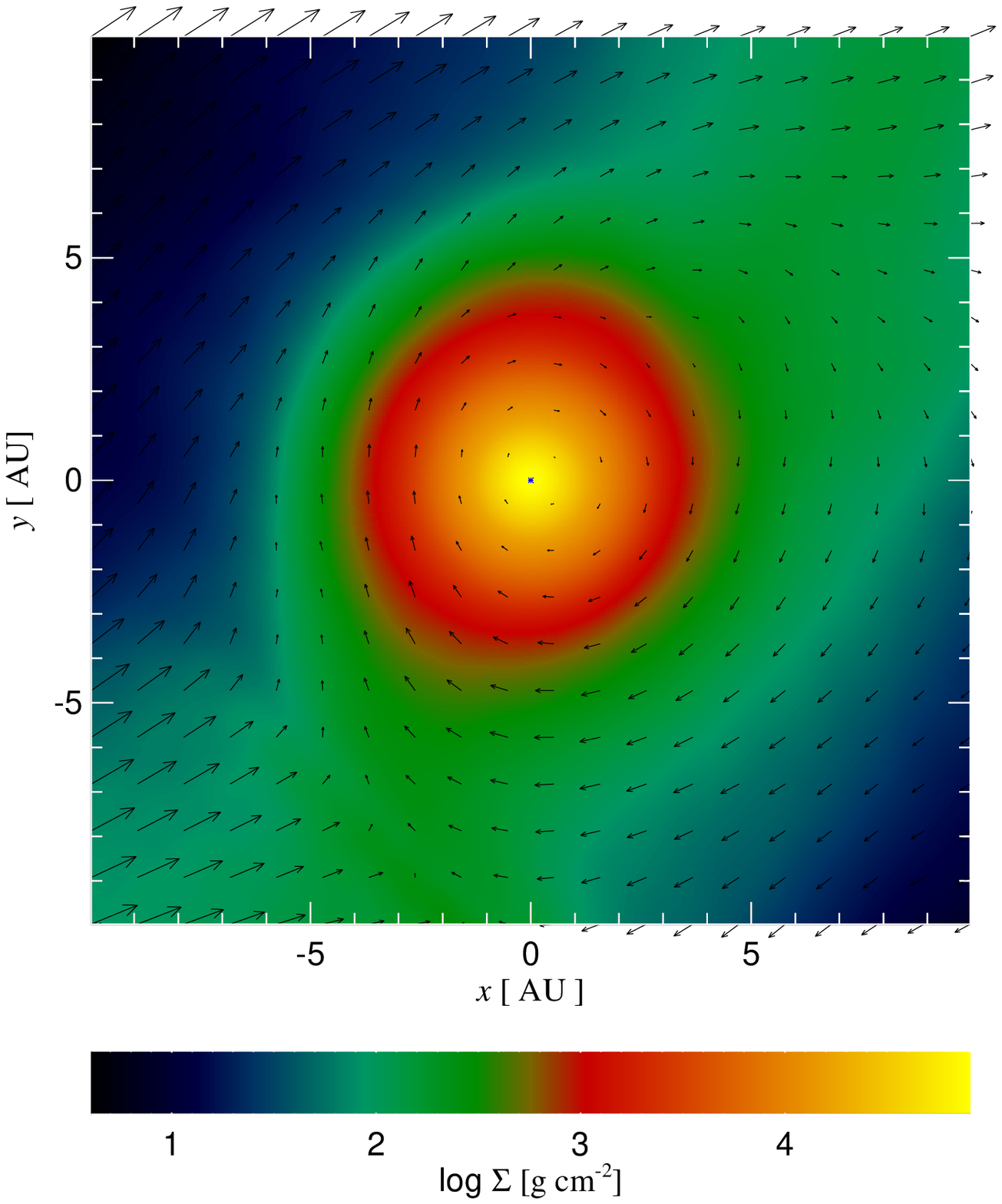,width=0.5\textwidth,angle=0}
\psfig{file=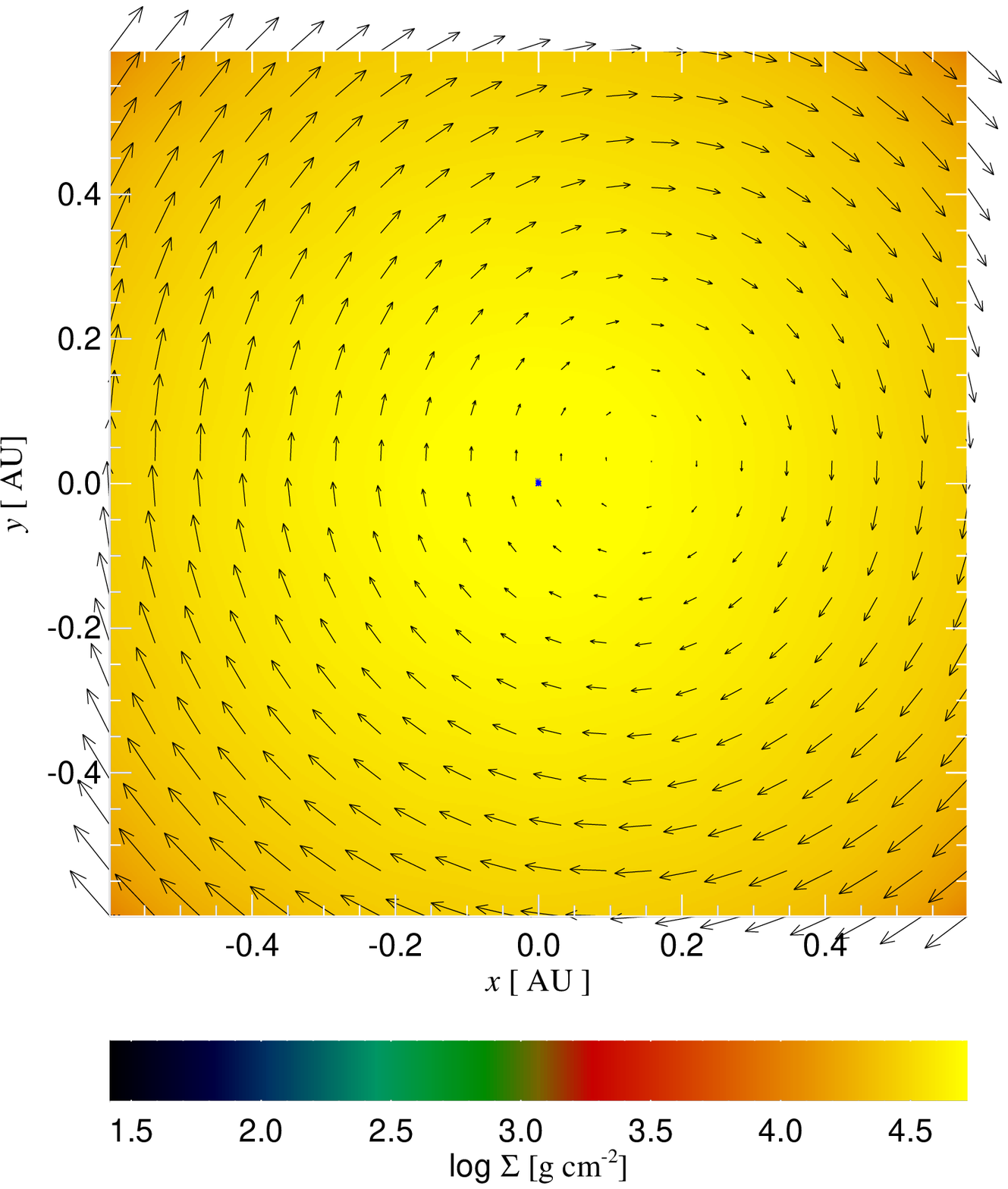,width=0.5\textwidth,angle=0}}
\caption{Top projections of the embryo closest to the star (located about 40
  AU in the south-west direction in the figure). Left: extended view, showing
  that the embryo is disturbed by the interactions with the parent star and other
  embryos. Right: zoom in on the same
  embryo. The figure is centred on the ``super-Earth''
  core \citep[see][]{ChaNayakshin10}. Note the offset between the gas envelope's
  centre of rotation and the super-Earth location.}
\label{fig:disturbed_embryo}
\end{figure*}

Our final example of rotating embryos found in simulations is shown in Figure
\ref{fig:disturbed_embryo} which shows the face-on projection of an embryo
closest to the star in the left panel of figure 4 of
\cite{ChaNayakshin10}. The embryo is within slightly less than 40 AU from the
parent star (located south-west in the figure) and has just interacted with
another embryo outside of the figure and located north-west. This is an
example of an embryo significantly perturbed by the tidal field of the star
and other interactions in the disc. The central dot in the left panel of the
figure is the super-Earth solid core (we do not show grains smaller than 100
cm in this figure). The right panel of figure \ref{fig:disturbed_embryo} shows
the central part of this embryo, centred on the solid core. Note that the
rotation pattern of the gas component is offset by about $0.15$ AU from the
solid core in this case.

The spin axis of this embryo is more strongly inclined away from the disc axis
of symmetry; the inclination angle for the embryo is slightly larger than
$30^\circ$. This is very likely to be the result of at least two interactions
that the embryo have had earlier. In particular, the embryo had merged with
another smaller one, and there were a close passage of a massive embryo
\citep[see][]{ChaNayakshin10}.

\section{Natal spin of rocky planets}\label{sec:rocky_spin}

We now assume that the embryos are roughly in a solid body rotation with
angular frequency $\Omega_0 = \xi_{\rm rot} \Omega_{\rm break}$, where $0 <
\xi_{\rm rot} < 1$ is a parameter. From our numerical simulations to date, and
also \cite{BoleyEtal10}, $\xi_{\rm rot} \sim 0.1$. $\xi_{\rm rot} =1$
corresponds to the maximum rotation frequency.  We write the mass of the
embryo $M_{\rm emb} = 10 M_J m_1$ and use the fiducial parameters for the
model embryo from \cite{Nayakshin10c}, which yields $R_{\rm emb} =
0.8$~AU~$t_4^{-1/2}$, where $t_4 = t/10^4$ years is the age of the embryo. The
minimum rotation period of an embryo is
\begin{equation}
T_{\rm min,0} =  2\pi \;\Omega_{\rm break}^{-1} = 7 \;t_4^{-3/4} m_1^{-1/2}  \;\hbox{yrs}\;.
\label{trot_emb2}
\end{equation}
The minimum angular frequency is likely to be the angular frequency of disc
rotation at the location where the embryos are born. This corresponds to the
maximum spin period of
\begin{equation}
T_{\rm max,0} = 10^3 \; R_{\rm 100 AU}^{3/2}\;\hbox{yrs}\;.
\label{trot_emb1}
\end{equation}
Typical rotation period of an embryo is then $T_{\rm rot,0} \sim$ 100 years.

Now let a rocky planet of mass $M_p$ be formed by a direct gravitational
collapse of a solid-dominated region, such as the ``grain cluster''
\citep{Nayakshin10a} (we first assume that rotational support
  is not important). The dust density, $\rho_d$, of such regions may be
higher than that of the gas by the factor of up to $f_g^{-1/2} \sim 10$, where
$f_g$ is the mass fraction of dust in the embryo. For typical opacity values
this implies the dust densities in the range of $10 ^{-9}$ to $10^{-7}$
g~cm$^{-3}$. The specific angular momentum of the collapsing fragment is $\sim
\xi_{\rm rot} \Omega_0 R_{\rm fr}^2$, where $R_{\rm fr} = (3 M_p/ 4\pi
\rho_d)^{1/3}$ is the linear dimension of the fragment.

The specific angular momentum of a rocky planet at birth is that of the
material that made it, and thus the rotation rate of the rocky planet is
\begin{equation}
T_p = T_{\rm rot,0} \left(\frac{\rho_d}{\rho_p}\right)^{2/3}\;.
\label{trot_p1}
\end{equation}
As $T_{\rm min,0} < T_{\rm rot,0} < T_{\rm max,0}$,
we arrive at the following estimate of the natal rotation period of rocky
planets:
\begin{equation}
0.25 \;t_4^{-3/4} m_1^{-1/2} \left(\frac{\rho_{-8}}{\rho_p}\right)^{2/3} < T_p <
41 \; R_{\rm 100
  AU}^{3/2}
\left(\frac{\rho_{-8}}{\rho_p}\right)^{2/3} \;,
\label{t_bounds}
\end{equation}
in hours, where $\rho_d = 10^{-8} \rho_{-8} $ g cm$^{-3}$. Alternatively, writing
$T_{\rm rot,0} = T_{\rm min,0} \; \xi_{\rm rot}^{-1}$,
\begin{equation}
T_p = 2.5 \;\frac{0.1}{\xi_{\rm rot}}\;t_4^{-3/4} m_1^{-1/2}
\left(\frac{\rho_{-8}}{\rho_p}\right)^{2/3} \;.
\label{t_xi}
\end{equation}

Now the shortest possible rotation period for a planet of density $\rho_p$ is
actually
\begin{equation}
T_{\rm break} = 2\pi (3/4\pi G \rho_p)^{1/2} \approx 1.5 \;\hbox{hours }
\label{t_br_p}
\end{equation}
for $\rho_p = 5$~g~cm$^{-3}$. This is longer than the minimum period estimated
in equation \ref{t_bounds}. This implies that for the most rapidly rotating
embryos the direct gravitational collapse formation route for rocky planets is
limited by the angular momentum effects. The initial configuration of rocky
planets may thus be oblate spheroids. Even more rapid rotation may cause
ragmentation and formation of binaries (see \S \ref{sec:Moon}) or multiples
inside the gaseous embryo.

So far we considered formation of the solid core by a single gravitational
collapse episode of a large dust reservoir. If further core growth continues
by accretion of small (cm-sized) solids well coupled to the gas \citep[as in
  the 1D models of][]{Nayakshin10b}, then that process could slow down the
planet's rotation. If the core's growth is dominated by collisional impacts of
larger solids, say $10$ km or more, then these may be shown to be decoupled
from the gas, as the gas-solid friction is unimportant for such large
objects. The result -- a spin-up or spin-down -- would then depend on the
balance of the flows of solids striking the planet with prograde or retrograde
angular momentum \citep[as in ``standard theory'',
  see][]{Giuli68,Harris77}. In general we would expect further solid accretion
to slow down the initially high spin of rocky planets.

\section{Rotation of giant planets}\label{sec:giant_rotation}

We now assume that for gas giant planets the solid core's mass is small
compared with the total planet's mass. If core's mass is a significant
contributor then the arguments of the previous section apply and hence the
planet is likely to be rapidly rotating at birth.

The maximum specific angular momentum of gas accreting onto the growing giant
planet of mass $M_p$ can be estimated as $J_p \sim \Omega_0 R_A^2$, where
$\Omega_0$ is the angular frequency of the giant planet's embryo, and $R_A$ is
the accretion radius defined by $R_A = 2 GM_p/c_s^2$, where $c_s^2$ is the
sound speed in the embryo. This estimate would be correct if gas accretes onto
the planet in a \cite{Bondi52}-like manner, so that beyond $R_A$ the gas moves
radially inward subsonically, with gas density and pressure deviating little
from its values at infinity, whereas gas inside the accretion radius plunges
on the accretor in a free fall. If the flow is deterred by angular momentum
barrier at intermediate radii or by thermal effects due to energy release by
the planet, then viscous torques are likely to carry some angular momentum
away as in the accretion disc theory of \citep{Shakura73}.

Accordingly,  we estimate that the minimum spin period of gas giants at birth is
\begin{equation}
T_{\rm giant} \ge T_{\rm rot,0} \frac{R_p^2 c_s^4}{(2GM_p)^2} = 3.3\; T_3^2
m_p^{-4/3}\rho_p^{-2/3} \frac{T_{\rm rot,0}}{100\;\hbox{yrs}} \;.
\end{equation}
where $T_3$ is the embryo's temperature in $10^3$ K, $T_{\rm rot,0}$ is the
embryo's rotation period, $m_p$ is the planet's mass in Jupiter masses, and
$\rho_p$ is the planet's mean density. The fastest rotation is again
comparable to the the break-up period (for Jupiter, $T_{\rm br} \sim 2.9$
hours). Hence the gas accretion estimate also predicts a rather fast initial
rotation, although with a range of uncertainties possible due to a non-linear
physics of envelope accretion \citep[e.g.,][]{PollackEtal96}.

\section{Constraints from the Solar System}\label{sec:obs}

\subsection{Rotation of planets}\label{sec:planet_spin}

Five out of nine Solar System planets rotate rapidly in the prograde fashion,
that is, in the direction of their revolution around the Sun (the Sun spins in
the same direction too). The exceptions are: the two inner rocky planets,
Mercury and Venus, whose spins are thought to have been strongly affected by
the Sun; Pluto, the dwarf planet with a weight of just a fifth of the Moon;
and Uranus, whose orbit is inclined at more than $90^\circ$ to the Sun's
rotational axis. Therefore, out of the major eight planets not strongly
affected by the Solar tides, the only exception to the prograde rotation is
Uranus. With exception of the two inner planets, the rest spin with a period
of between about half a day and a day. These spin rates are large: increasing
them by a factor of a few to a few ten would tear the planets apart by
centrifugal forces.

The origin of these large and coherent planetary spins is difficult to
understand \citep[e.g.,][]{DonesTremaine93} in the context of the
``classical'' Earth assembly model \citep[e.g.,][]{Wetherill90}. In this model
rocky planets grow by accretion of smaller rocky fragments. As terrestrial
planets are physically very small, e.g., $R\simlt 10^9$ cm, compared with
dimensions of the disc (their orbits), one expects that the accretor should
receive nearly equal amounts of positive and negative angular momentum. The
final spin is a result of a delicate cancellation of these positive and
negative angular momentum impacts. Not surprisingly, the result is highly
sensitive to the assumptions about the orbits of the bodies accreting on the
proto-planet \cite{Giuli68,Harris77}. For this reason, the large spins of
Earth and Mars are most naturally explained by one or a few ``giant''
planetesimal impacts \citep[][]{DonesTremaine93}. However,
  such impacts would have to be very special to give the Earth and the Mars
  similar spin directions also closely matching that of the Sun.

In the context of the tidal downsizing scenario, as we argued above, most of
the planets would rotate coherently, with exception of those whose parent
embryos have undergone direct collisions or close passages of other
embryos. Therefore the rotation rates and directions of the Earth and the Mars
would be the norm rather than exception in this picture. The rotation pattern
of gas giants in the Solar System may be explained in a similar fashion but
due to accretion of gas onto the solid cores (see \S
\ref{sec:giant_rotation}).

\subsection{The origin of the Moon}\label{sec:Moon}

The subject of planetary rotation closely sides with that of formation of
planetary satellites. By the ratio of the mass to that of the primary planet,
the Moon is the heaviest satellite amongst the major Solar System planets.
The Moon is generally believed to have been formed due to a giant impact of a
large solid body on the Earth \citep{HartmannDavis75,CA01}. Numerical
simulations of giant impacts indicate that the Moon would have been mainly
\citep[$\sim$ 80\%, see][]{Canup08} made of the impactor (named Theia).

Various composition measurements indicate that the mantle and the crust of the
Earth and the Moon are very similar \citep[see a list of references
  in][]{dMvW09}. In fact, even the oxygen isotope composition
  of the two bodies were measured to be similar which initially appeared to be
  consistent with the idea of Theia forming very nearby to Earth
  \citep{WiechertEtal01}. However, more recent oxygen isotope ratio
  measurements of lunar samples reveal an oxygen isotope composition that is
  not just similar but is indistinguishable from the terrestrial samples, and
  clearly distinct from meteorites coming from Mars and Vesta, which motivated
  suggestions of complicated and highly efficient mixing processes during the
  hypothesised Earth-Theia collision \citep{PahlevanStevenson07}. 

We shall now consider these observational facts in the context of the tidal
downsizing scenario. A simple thought is this: perhaps Theia, the body that
struck the proto-Earth, could have been formed inside the same giant planet
embryo that formed the Earth, explaining the compositional similarities.

In addition, tidal downsizing picture also allows for a more optimistic look
at the the famous fission hypothesis by George Howard Darwin \citep[see, e.g.,
][]{Binder74}. There is not enough angular momentum for the proto-Earth for
fission to occur \citep[e.g., for a review see][]{BossPeale86} in the standard
planetesimal accretion theory. The tidal downsizing hypothesis does have
enough angular momentum, as we argued in \S \ref{sec:rocky_spin}. On the other
hand, we calculate that the natal angular momentum of the Earth rotating near
the break up limit would be about a factor of 3 higher than the present day
value in the system. A mechanism to get rid of the angular momentum would then
be required in this picture. Assuming that the excess angular momentum is
carried away by rocky bodies or gas with angular momentum (with respect to the
Earth) equal to that at the Hill's radius of the Earth would require as much
as $1/3$ Earth masses of material. This is however not large with respect to
the original giant envelope mass, and thus should not be ruled out outright.

Finally, fast rotation of the ``grain cluster'' of solids
\citep[see][]{Nayakshin10a} may prevent its condensation into a single
body. Referring back to figure \ref{fig:disturbed_embryo} where the
super-Earth is offset from the rotational centre of the gas embryo, it is not
impossible that in such a case a second core is born in the rotational centre,
creating a binary solid-core system.  Further work is needed to test which of
the three points of view, if any, could explain the data, but it seems clear
that forming the parent bodies of the Earth and the Moon inside the same
embryo would help to explain the extraordinary degree of oxygen isotopes
similarity between them.

\subsection{Other constraints from the Solar System}\label{sec:other}

The referee of this paper has pointed out that tidal downsizing hypothesis
does not currently have explanations for at least the following observations
of the Solar System: the Late Heavy Bombardment event, the exact chemical
composition of the rocky inner planets, and the orbital evolution of the outer
giant planets. These issues are outside of the scope of this paper, but we
hope to clarify them in our future work.

\section{Conclusions}

In this short article we estimated the rotation rate of planets, both
terrestrial and giant, in the context of the tidal downsizing hypothesis for
planet formation. We showed that such planets could potentially be rotating
near their break up limit at formation, although there are mechanisms for
lowering the initial spins. The default direction of the spins coincides with
that of the parent disc. This may explain the fast and coherent rotation
pattern of most of the Solar System planets. Exceptions to the coherent
rotation may be due to embryo-embryo interactions that appear to occur
frequently in the simulations. We also argued that the Moon could have formed
inside the same parent embryo as the Earth, explaining compositional
similarities between the two bodies.

\section{Acknowledgments}

Theoretical astrophysics research at the University of Leicester is
supported by a STFC Rolling grant. The authors thanks Seung-Hoon Cha for his
permission to use the simulation data of \cite{ChaNayakshin10} to illustrate
the analytical theory presented here. The author thanks the
  anonymous referee for a useful report that made the shortcomings of the
  tidal downsizing hypothesis clearer.


\label{lastpage}

\end{document}